\documentclass[aps,pre,amsmath,amssymb,lengthcheck,showpacs,superscriptaddress]{revtex4-1}
\usepackage{graphicx}
\usepackage{color}
\usepackage[version=3]{mhchem}
\begin{document}
\title{Water sub-diffusion in membranes for fuel cells}
\author{Quentin Berrod}
\affiliation{LLB, CEA, CNRS, Universit\'e Paris-Saclay, CEA Saclay 91191, Gif-sur-Yvette, France}
\affiliation{Lawrence Berkeley National Laboratory, Energy Storage Group, 94720 Berkeley, USA}
\author{Samuel Hanot}
\affiliation{Institut Laue-Langevin - 71 Avenue des Martyrs - CS 20156 - 38042 Grenoble}
\affiliation{Unit\'e de Bioinformatique Structurale, Institut Pasteur, Paris, France}
\affiliation{UMR 3528, CNRS, Paris, France}
\author{Armel Guillermo}
\affiliation{Univ. Grenoble Alpes, CEA, CNRS, INAC, SYMMES, F-38000 Grenoble}
\author{Stefano Mossa}
\email{stefano.mossa@cea.fr}
\affiliation{Univ. Grenoble Alpes, CEA, CNRS, INAC, SYMMES, F-38000 Grenoble}
\author{Sandrine Lyonnard}
\email{sandrine.lyonnard@cea.fr}
\affiliation{Univ. Grenoble Alpes, CEA, CNRS, INAC, SYMMES, F-38000 Grenoble}
\date{\today}
\begin{abstract}
We investigate the dynamics of water confined in soft ionic nano-assemblies, an issue critical for a general understanding of the multi-scale structure-function interplay in advanced materials. We focus in particular on hydrated perfluoro-sulfonic acid compounds employed as electrolytes in fuel cells. These materials form phase-separated morphologies that show outstanding proton-conducting properties, directly related to the state and dynamics of the absorbed water. We have quantified water motion and ion transport by combining Quasi Elastic Neutron Scattering, Pulsed Field Gradient Nuclear Magnetic Resonance, and Molecular Dynamics computer simulation. Effective water and ion diffusion coefficients have been determined together with their variation upon hydration at the relevant atomic, nanoscopic and macroscopic scales, providing a complete picture of transport. We demonstrate that confinement at the nanoscale and direct interaction with the charged interfaces produce anomalous sub-diffusion, due to a heterogeneous space-dependent dynamics within the ionic nanochannels. This is irrespective of the details of the chemistry of the hydrophobic confining matrix, confirming the statistical significance of our conclusions. Our findings turn out to indicate interesting connections and possibilities of cross-fertilization with other domains, including biophysics. They also establish fruitful correspondences with advanced topics in statistical mechanics, resulting in new possibilities for the analysis of Neutron scattering data.
\end{abstract}
\maketitle
\section{Introduction}
\label{sect:intro} 
The dynamical nature of confined water determines the structure-to-function relationship in supra-molecular hydrated assemblies. For instance, water anomalous transport at extended interfaces, or under severe confinement conditions, plays a key role in mediating performance and function of {\em natural} assemblies, as proteins.~\cite{bellissent2016water} Similarly, deviations from the bulk behavior alter the functional properties of soft confining {\em synthetic} materials, employed in a wide range of applications in nanotechnology and environmental or energy science. The entanglement of spatial confinement and host-fluid interactions determines the mobility of fluid molecules within ionic nanostructures, controlling charge transfer and transport. The identification of the diffusion mechanisms of ions and water confined in soft charged media is a general goal of materials science, and can guide the optimization of ionic nanoorganized materials with selective ion-driven functionalities. 

Proton-conducting hydrated ionomers are a class of soft ionic self-assembled systems, thoroughly scrutinized for their potential as separators in energy conversion devices, including fuel cells, electrolysers, and red-flow cells. In particular, ionomers exhibit multi-scale hydration-dependent morphologies, which include a mechanically robust polymer host matrix containing randomly distributed irregular ionic nanodomains. Proton conductivity in these structures depends on both the macroscopic total amount of water adsorbed in the network of interconnected channels, and the microscopic structure of the fluid molecules. It has been proposed that efficient proton transfer originates from a hydration-dependent balance between vehicular (mass) transport of hydronium ions and hydrogen bonds driven structural (charge) diffusion.~\cite{kreuer2004transp-rev} Both processes are affected by the local and long-range topologies of the ionic domains, and by the features of the polymer/water interface. 

Significant effort has already been spent in establishing the driving forces which dictate the ionomer performances, according to chemical architecture or processing conditions. Perfluoro-sulfonic acid (PFSA) compounds, as Nafion by Dupont, Aquivion by Solvay, Gore or 3M materials, have been particularly explored. These are made of a highly hydrophobic Teflon-like backbone, with grafted perfluorinated side-chains terminated by acidic functions.~\cite{mauritz2004state} For their well-established capability to form phase-separated morphologies inducing outstanding proton conductivities, they are the state-of-the-art fuel cell electrolytes. Despite their generalized use in applications, however, PFSAs do not reach sufficient proton conductivities in high temperature and low hydration conditions, are based on toxic chemistry, and imply high production costs. Further optimization or, as an option, development of alternative chemical architectures, remain crucial issues for a large-scale deployment of fuel cells.~\cite{hickner2004review-memb}

The rational design of advanced ionomers require to define criteria based on the structure-transport correlations. With this objective, it is critical to identify the relevant length (and, consequently, time) scales over which the hydration-dependent processes are active, and their mutual interplay. These cover a vast spectrum, encompassing molecular (host-fluid interactions), nanoscopic (hydrophilic/hydrophobic phase-separation, confinement, interfaces), mesoscopic (connectivity, topology, tortuosity), micrometric (grain organisation, grain boundaries) and macroscopic (conductivity) domains. 

Advanced spectroscopy and scattering techniques are powerful tools in this direction,~\cite{gebel2005tools,lyonnard2012neutrons} since they allow to explore structure~\cite{gierke1981,Rubatat2002,choi2005sorpt,lee2009process,liu2010ew,park2011oriented,kreuer2013flat,negro2013acidic-sites} and dynamics~\cite{Perrin2007,page2009relax,hallinan2010nonFick,luo2011iec,perrin2012temp,dalla2014IR,page2014nse,Berrod2015qens,Berrod2015} of matter at those scales. This effort has been efficiently integrated by numerical work, employing techniques such as ab-initio calculations~\cite{paddison2001-ioni, dupuis2010, paddison2013proton, vartak2013collective}, Molecular Dynamics (MD)~\cite{devanathan2007,devanathan2007struct,venkatnathan2007temp,devanathan2012struct,malek2008nanophase,liu2010ew,voth2012evb,daly2013benziger,savage2014persistent,Hanot2015,Hanot2016,arntsen2016simulation,savage2016proton,ghelichi2016ionomer} and DPD~\cite{wu2008Aq-Naf,wu2009nano-EW} simulations, and continuum approaches~\cite{weber2014critical,liu2015mesoscale}. Insight into properties and elementary processes of PFSAs as structure~\cite{Rubatat2002,devanathan2012struct,kreuer2013flat,Berrod2015}, ionization~\cite{paddison2001-ioni,dupuis2010,dalla2014IR}, sorption~\cite{choi2005sorpt,eikerling2009book,daly2013benziger}, proton conduction~\cite{eikerling2001prot,kreuer2004transp-rev,vartak2013collective,paddison2013proton,savage2014persistent,liu2015mesoscale,savage2016proton,arntsen2016simulation}, water diffusion~\cite{Perrin2007,devanathan2007,hallinan2010nonFick,damasceno2013inhomogeneous,savage2014persistent,damasceno2014morphology,Hanot2016}, polymer relaxation~\cite{page2009relax,page2014nse}, has been accumulated. The impact of the nature of polymer backbone~\cite{hickner2004review-memb,smitha2005rev-memb,ghelichi2016ionomer} and ionic side-chains~\cite{wu2008Aq-Naf, luo2011iec, ghelichi2016ionomer}, the nature of acidic sites~\cite{negro2013acidic-sites}, the density of charge~\cite{wu2009nano-EW,liu2010ew,luo2011iec}, elaboration methods~\cite{lee2009process}, and other factors~\cite{venkatnathan2007temp,park2011oriented,perrin2012temp,hiesgen2012afm} have also been studied. All these aspects define a space of possibly relevant parameters which is too extended for being exhaustively explored. As a consequence, a comprehensive approach directed towards a general (universal) understanding of fundamental mechanisms is still lacking.

Based on these considerations, we propose here an integrated experiments/simulations investigation of the structure-transport interplay of perfluoro-sulfonated compounds. Our goal is to unambiguously identify an universal physical picture for transport of water absorbed in PFSA, irrespective to the details of the chemistry, and in the spirit of a fundamental general understanding of the entire class of materials. We have considered the benchmark membrane Nafion (long side-chain), the more recent competitor Aquivion (short side-chain), and the perfluoro-octane-sulfonic acid (PFOS) ionic surfactant. This latter is a macro-molecule with a structure very similar to that of the side-chain of Nafion, and has been demonstrated to be an efficient alternative for clarifying the local structure of the ionomer.~\cite{Lyonnard2010,Berrod2015,Hanot2015} We have performed an extensive multi-scale experimental characterization of Aquivion, and Molecular Dynamics simulations of Nafion and Aquivion. These extended new data sets integrate and complete the range of experimental data we have put together recently,~\cite{Berrod2015} and simulation results on ionic surfactants.~\cite{Hanot2015,Hanot2016} We establish that {\em sub-diffusion} of adsorbed water is the key universal property lying at the bottom of the multi-scale transport behavior within the PFSA materials. Surprisingly, this feature has been until now somehow overlooked in the present context (with only a few exceptions, see below), while it is a lively subject of interest in biophysics, among other fields. We believe we are in the position, for the first time to the best of our knowledge, to propose a unique coherent vision of the structure/function interplay dictating the main properties of PFSA materials, at all length scales.
\begin{figure*}[h!t]
\centering
\includegraphics[width=1\linewidth]{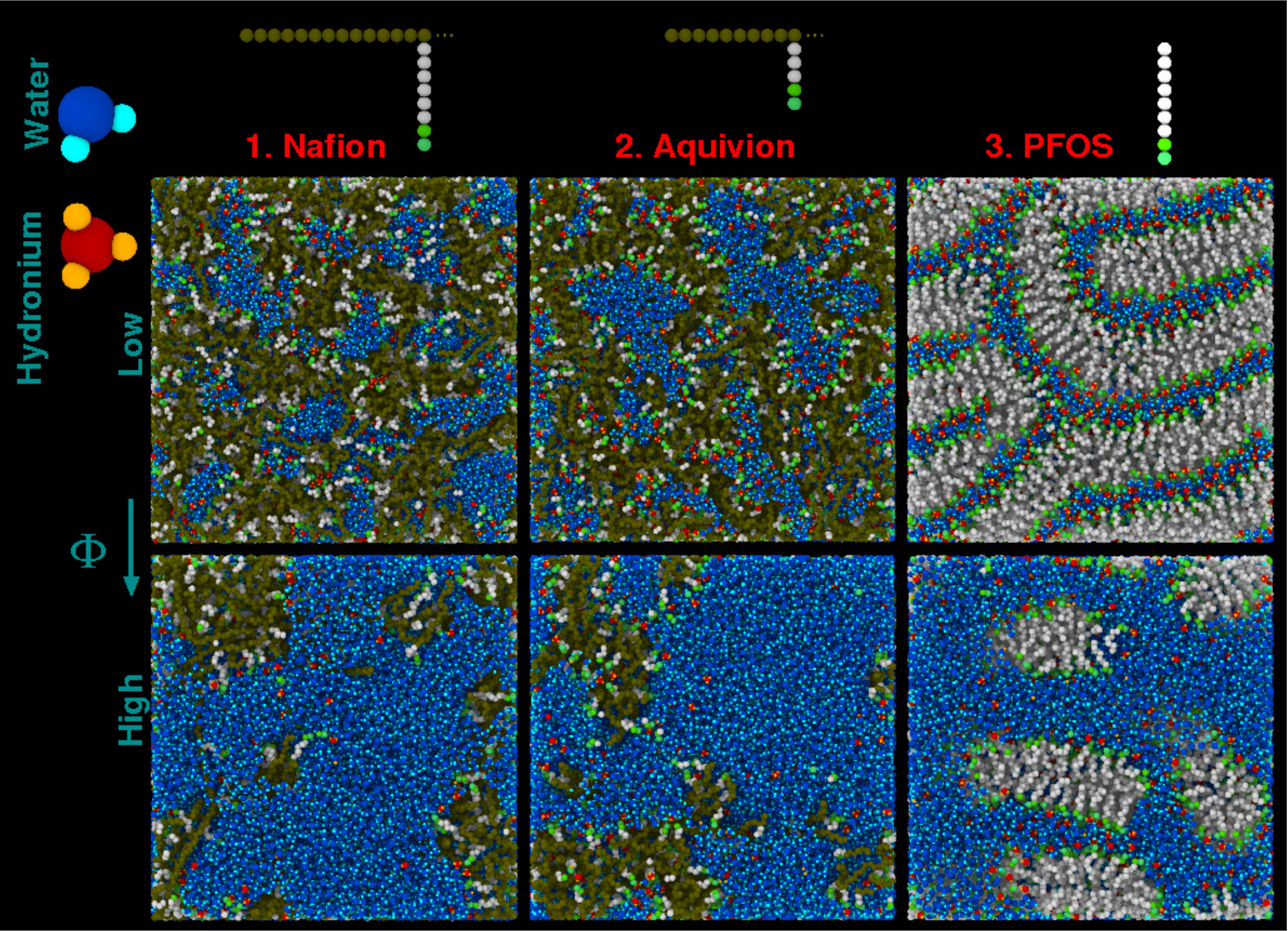}
\caption{
Sketches of the molecular models used in the MD simulations, together with typical snapshots of the considered systems at different levels of hydration. Water molecules~\cite{berendsen1987missing} and the hydronium ions~\cite{kusaka1998binary} are described with all-atoms resolution. Ionomers and ionic surfactants are modeled at a coarse-grained level,~\cite{Hanot2015} by using the indicated units as building blocks. We have represented in brown and white hydrophobic beads which, even if belonging to the same species, pertain to the polymer backbones or the side chains, to highlight more precisely the mutual organization of the structural features in the different materials. Two green beads schematize the \ce{SO3-} groups, water is in blue, hydronium complexes in red. We show typical self-organized morphologies for all the investigated materials, generated at low and high water volume fractions $\Phi\simeq$ 0.1 and 0.4, respectively. Colors in the snapshots are the same as in the sketches.
}
\label{fig:1}
\end{figure*}
\section{RESULTS AND DISCUSSION}
\label{sect:results}
In the following, we first underline the variations of morphology of the PFSA membranes and the model case of PFOS as obtained both by SAXS and MD, in different hydration conditions. Next, we provide a complete description of the dynamics of the hydration protons and charge carriers, by simultaneously discussing QENS, NMR and MD data for all materials. We finally focus on the hydronium ions dynamics, based on an integrated discussion of QENS and MD data. Details about the techniques used are given in the Methods section. Raw data and additional details on the analysis performed are reported in the Electronic Supporting Information (ESI$\dag$).  
\subsection{Quantification of the hydration-dependent nanoscale confinement}
\label{subsect:morphology}
Due to their amphiphilic character, hydrated membranes exhibit clearly separated hydrophobic and ionic domains. Although the microstructure of PFSAs has been studied since the early stage of ionomers development,~\cite{gierke1981} it continues to stimulate an impressive body of literature. In fact, details of the multi-scale structure of Nafion are still debated,~\cite{Rubatat2002,schmidt2008cyl,kreuer2013flat} due to the quite ill-defined features of the scattering spectra, non-uniqueness of the models employed for fitting the data, persisting difficulties in obtaining high-resolution images (although important information can come from modern 3-dim techniques~\cite{weber2014TEM}) and a very extended range of relevant scales to probe by numerical simulation with sufficient resolution. Still, there is a commonly accepted picture based on the organization of flat elongated polymer aggregates into a continuous ionic medium,~\cite{kreuer2013flat} whose extension increases with water content.~\cite{Rubatat2002,kreuer2013flat,fumagalli2015SANS, wu2008Aq-Naf, liu2010ew, wu2009nano-EW, malek2008nanophase, devanathan2007struct} Swelling is accompanied by important modifications of the topology of the interface between the ionic domains and the polymer matrix, to minimize the interfacial energy. This is a universal mechanism controlling, for instance, the lamellar-to-hexagonal transition observed in surfactants at intermediate water contents.~\cite{nallet2011,dubois2000tensio} The morphology and swelling of PFSAs have been shown indeed to resemble that of ionic surfactants,~\cite{Berrod2015} assigning to side-chains and acidic functions the predominant effect in controlling size, shape and organization of the ionic domains. 

Salient {\em qualitative} features of the expected morphologies can already be grasped by visually inspecting in Figure~\ref{fig:1} snapshots of all materials, generated by MD at low ($\simeq 0.1$) and high ($\simeq 0.4$) water volume fraction, $\Phi$. At low $\Phi$, the ionic domains are intercalated between inter-connected elongated polymer aggregates, ordering in a neat lamellar phase in the PFOS. At high $\Phi$, the connectivity of the ionic domains is substantially higher, in agreement with other studies.~\cite{wu2008Aq-Naf,malek2008nanophase,devanathan2007struct} The structure is now composed of rather isolated hydrophobic objects embedded in the ionic phase, formed by percolating bulk-like water pools. In PFOS the formation of the micellar phase is very clear. Also, we recognize well-defined interfaces at the boundaries of hydrophobic and hydrophilic regions, a common feature of these materials. Side chains are projected outward from the polymer aggregates, with the sulfonate ions (in green) localized at the interfaces, where the hydronium ions (red) mainly condense at low $\Phi$. Also, note that at similar $\Phi$, the shorter the side-chain, the  thinner (on average) the polymer aggregates. 

A {\em quantitative} unified analysis of experimental and simulation data of all these materials imposes to be specific about two issues. First, we must be able to define the degree of hydration consistently for materials characterized by different mass and charge contents. In Figure~S1$\dag$, we plot the sorption isotherms for the ionomers. At given relative humidity, the {\em total} water uptake ($\Phi$) in Aquivion is higher than in Nafion due to lower \textcolor{black}{equivalent weight, EW (790 and 1100~g/eq, respectively)}. The {\em local} hydration levels $\lambda$ (number of water molecules per sulfonic acid group), in contrast, are identical and turn out to be the most adapted parameter when comparing different materials. Note that this choice is also the most indicated for simulations, where $\lambda$ is determined unambiguously (contrary to $\Phi$), by simply counting the number of water molecules. 

Second, in order to establish a precise measure of the transport/morphology correlation, the latter must be concisely encoded in a single descriptor determinable in experiments. Surprisingly, the mere average size of the ionic domains, $d_w$, turns out to be sufficient for this aim, in particular in hydration conditions where confinement at the nanoscale, more than the total water content, primarily determines the dynamics.~\cite{Berrod2015} Measurements of $d_w$ can be determined by SAXS experiments, that we have performed on Aquivion membranes to complement those on Nafion and PFOS.~\cite{Rubatat2002,Lyonnard2010} We show in Figure~S2~a)$\dag$ the 1-dim spectra for Aquivion, at the indicated values of $\lambda$. The regular organization of the ionic domains produces a typical correlation feature, the ionomer peak, whose position, $Q_{\mathrm{iono}}$, provides the mean separation distance between neighboring polymer aggregates, $d_{\mathrm{iono}} = 2\pi/Q_{\mathrm{iono}}$ (Figure~S2~b). Note that, at a given $\lambda$, this distance is on average larger in Nafion than in Aquivion, as a direct consequence of the reduced length of side-chains in the latter, providing a quantitative confirmation of the inspection of Figure~\ref{fig:1}. \textcolor{black}{From $d_{\mathrm{iono}}$ we can finally obtain $d_w = d_{\mathrm{iono}} - d_0$, where $d_0$ is the mean size of the polymer aggregates deduced from the extrapolation $\lambda\rightarrow 0$ of the swelling laws (Figure~S2~b)$\dag$. Analogous procedures are followed for simulation data.~\cite{Hanot2015}}
\begin{figure}[t]
\centering
\includegraphics[width=0.45\textwidth]{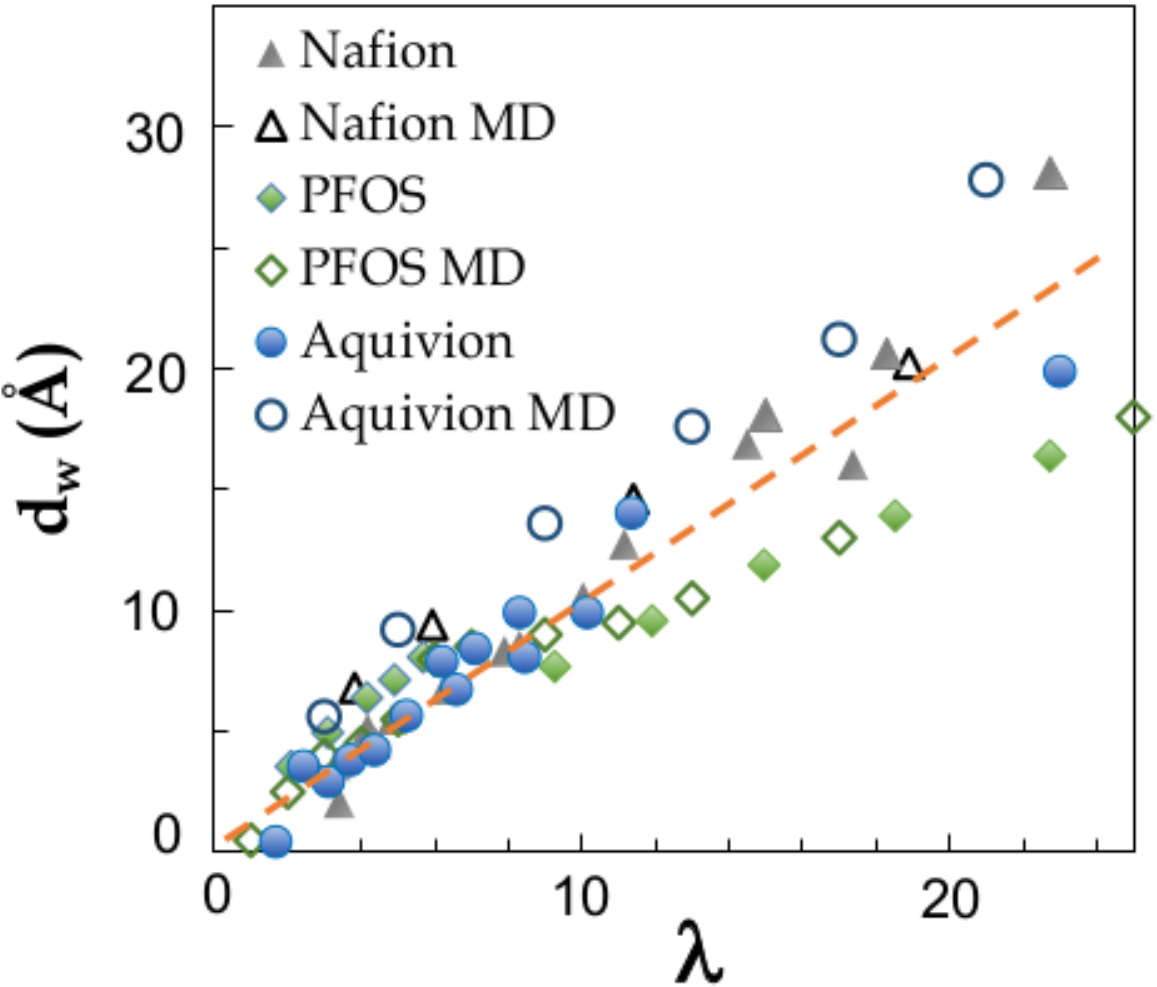}
\caption{
Average size of the ionic domains, $d_w$, calculated as detailed in the text, as a function of the hydration level, $\lambda$. We show together data sets obtained by SAXS/SANS and MD simulation for Aquivion, Nafion, and PFOS. \cite{Berrod2015} \textcolor{black}{The dashed line $d_w \propto \lambda$}, is a guide for the eyes. \textcolor{black}{Analogous procedures are used to calculate $d_w$ for both data sets. A complete discussion of these data is included in the main text.}
}
\label{fig:2}
\end{figure}

Fully consistent experimental and MD $d_w$ data for the ionomers and surfactants are all shown as a function of $\lambda$ in Figure~\ref{fig:2}. For all materials $d_w$ monotonously increases with $\lambda$. More precisely, we clearly observe that the ionomers data stay very close at all hydration, \textcolor{black}{exhibiting a linear variation $d_w(\lambda) \propto \lambda$ of slope 1 for $d_w<$~10~\AA} ~and $\lambda$-values pertaining to the affinely swelling lamellar phase (dashed line). This evidence indicates that differences in the polymer backbone structure, including \textcolor{black}{EW} or side-chains length, do not primarily control the features of the nano-confinement. In addition, in the same affine deformation region the swelling behavior in PFOS is completely analogous to that of the ionomers. The elongated polymer aggregates therefore seem to be diluted very similarly to the surfactant lamellar phases, in the same hydration region where the latter are stable. We conclude from these data that the local topology of the ionic domains in all PFSA materials is ultimately determined by the presence of strongly hydrophobic macromolecular sections and the superacidity of the sulfonic-head groups. The local organization of flat hydrophobic objects and the neat hydrophilic/hydrophobic interfaces are, therefore, general properties which are not sensitive to variations of the density of charge, backbone design, or even complete absence of the latter. In addition, simulation seems indeed able to provide synthetic structures which we can exploit as templates for real systems.
\begin{figure*}[t]
\centering
\includegraphics[width=0.7\textwidth]{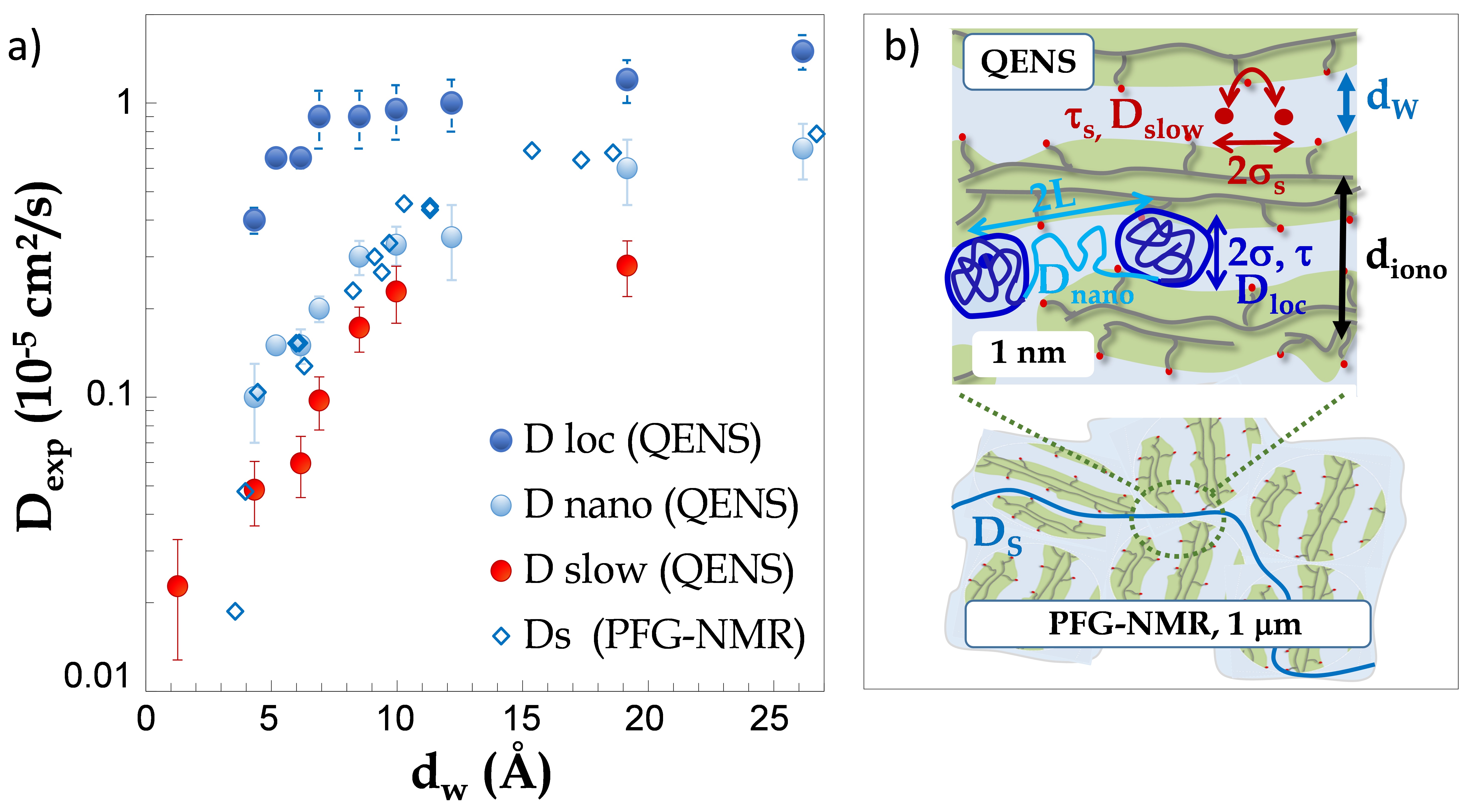}
\caption{
a) Diffusion coefficients extracted by the analysis of the data from QENS ($D_\text{loc}$, $D_\text{nano}$ and $D_\text{slow}$), and PFG-NMR ($D_S$) for Aquivion, as a function of the mean size of ionic domains $d_w$. These data are discussed at length in the main text. For the details of data analysis and modeling we refer the reader to the ESI$\dag$. \textcolor{black}{b) Schematic representation of fast (diffusive, blue) and slow (hopping, red) protons and their diffusion path-ways at local, nano and mesoscales. The dynamical parameters (characteristic times and distances) used for the rationalization of experimental QENS and PFG-NMR data are also indicated, together with the structural sizes. $D_{\text{loc}}$, $D_{\text{slow}}$, $D_{\text{nano}}$, and $D_S$ are measured at the $\AA$~/~1-10 ps, $\AA$~/~100 ps, nm~/~1 ns, $\mu$m~/~ms length and time scales, respectively.}
}
\label{fig:3}
\end{figure*}

\subsection{Multi-scale dynamics}
\label{subsect:water_diff}
The mobility of protons in PFSAs compounds can be studied by combining multi-resolution QENS and PFG-NMR measurements. QENS, in particular, is the only technique which can probe molecular-level dynamics at the ps to ns time scales, on length scales of a few \AA~$\,$ to 1~nm. This feature is invaluable for underlying the modifications of molecular mobility upon confinement at the nano-scale, in particular in hydrogen-containing systems.~\cite{berrod2016IL,mamontov2008qens,michot2007qens,zanotti1999qens} Quantitative information on the nature of the motions in terms of relaxation times, characteristic distances and diffusion coefficients, can be extracted by suitably modeling the incoherent dynamic structure factor, $S(Q,\omega)$. Multi-resolution QENS is often needed for complex systems whose dynamics span a wide temporal range.~\cite{Perrin2007,bedouret2014,berrod2016IL,ferdeghini2016IL} In this case, sophisticated multi-scale models that integrate the various processes must be employed to consistently reproduce the experimental spectra. These procedures are annoyed by two problematic aspects. First, \textcolor{black} {Neutron scattering (similarly to NMR)} cannot a-priori discriminate among bare charge carriers (\ce{H+}) and the hydration protons (\ce{H2O}). As a consequence, the total signal is the sum of various contributions, making modeling extremely difficult. Second, even if the probed $(Q,\omega)$-domain is quite extended, data analysis ultimately condenses the information in a restricted set of spectroscopic parameters, corresponding to only a few separated relevant length and time scales. How to overcome this difficulty is one of the main messages of this work (see below).
 
We have recorded QENS spectra \textcolor{black} {of hydrated Aquivion membranes} at five energy resolutions, and have analyzed the data following a method used recently for Nafion.~\cite{Perrin2007} We refer the reader to the Methods section and the ESI$\dag$ for all details, and focus our discussion on the determination of the relevant diffusion coefficients. Data analysis provides convincing evidence of the existence of two varieties of protons with distinct dynamics, also see Ref.~\cite{goossens2016unlocking}. (Technically, this is established by the need to consider two separate quasi-elastic terms in the $S(Q,\omega)$, see the ESI$\dag$.) These populations, are dubbed as {\em fast} and {\em slow}, with typical correlation times of 1 to 10~ps and hundreds of ps, respectively. More precisely, fast protons have a diffusive behavior (the associated quasi-elastic component, $\Gamma(Q)$, broadens continuously with $Q$), while the slow ones undergo localized motions ($\Gamma(Q)$ is $Q$-independent). These findings are analogous to those for Nafion~\cite{Perrin2007} and surfactants.~\cite{Berrod2015qens,Lyonnard2010} The diffusive component can be further described in terms of the generalized Gaussian model for localized translational motion~\cite{Volino2006} (GMLTM, see Methods section), while localized motions are modeled as hopping between two equivalent sites.~\cite{bee1988qens} The physical picture behind these frameworks is exemplified by the cartoons of \textcolor{black} {Figure~\ref{fig:3}~b) (also see Figures~S3$\dag$)}, raw data and fitting results are reported in Figures~S4-S10$\dag$. 

The fraction of fast and slow protons and the associated characteristic relaxation times ($\tau$ and $\tau_s$) are shown in Figure~S8$\dag$, together with the analogous data for Nafion.~\cite{Perrin2007} The analogies in the two materials are noticeable: while the number of diffusive entities correlates linearly with the hydration level, an approximately constant number (three) of protons is involved in the slow hopping, at all values of $\lambda$. \textcolor{black}{Therefore, slow protons pertain to species which are dynamically distinct (not exchangeable on the timescale of a ns) from the water molecules diffusing along the channels. Interestingly, the slow population grows only mildly with the total amount of adsorbed water and is surviving even in the highly swollen network of ionic channels (large bulk-like well-connected pools are formed at full hydration).} The raw data for $\tau_s$ and the associated slow protons hopping distance $2\sigma_s$ are shown in Figure~S9$\dag$ and are, again, very similar to those pertaining to Nafion. From these quantities, we can devise a parameter with the dimensions of a diffusion coefficient, $D_{\text{slow}}=\sigma_s^2 / \tau_s$. \textcolor{black}{(Insight into the nature of the highly localized protons and associated dynamical parameters} will be further discussed below.)

In the generalized GMLTM, protons pertaining to water molecules or hydronium complexes diffuse within soft droplets \textcolor{black}{characterized by the} average size $l=2 \sigma$. Diffusion is described as the result of random jumps between neighboring sites, with a mean jump-time $\tau$, and a local diffusion coefficient $D_{\text{loc}}$. In addition to this local (intra-droplet) mechanism, inter-droplets motions active on the nm length scale must be included in the analysis to account for the multi-resolution spectra (the Generalized Gaussian Model, see ESI$\dag$). This mechanism is described in the standard Fickian formalism of diffusion in a continuous medium, and quantified by the associated nano-scale diffusion coefficient, $D_{\text{nano}}$\textcolor{black}{, and the mean inter-droplet separation distance $2L$}. The raw data for $\tau$ and 2$\sigma$ are shown in Figure~S10$\dag$, where they are compared with the very similar values obtained for Nafion, at the investigated values of $\lambda$. In addition to the local and nanoscopic information extracted from QENS experiments, motions in the mesoscopic (continuum-like) range, {\em e.g.}, at time and length scales of the order of ms and $\mu$m, respectively, are explored by PFG-NMR experiments, where one evaluates the self-diffusion coefficient of water molecules, $D_S$, as detailed in the Methods section. 

\textcolor{black}{The dynamical mechanisms described above are pictured in the cartoons of Figure~\ref{fig:3}~b), where schematic representations of the fast (hopping) and slow (intra- and inter-droplet diffusion) proton motions are shown embedded within the nano- and meso-scale ionomer structures. Typical time and length-scales probed by QENS and NMR experiments are indicated, together with the various relevant dynamical parameters, e.g., characteristic times, distances, and diffusion coefficients.} The multi-scale analysis of the dynamics realized by combining multi-resolution QENS and PFG-NMR ultimately distills all information in a set of four diffusion coefficients, either determined directly or assembled combining values of length and time scales associated to different mechanisms. In Figure~\ref{fig:3}~a), we represent the variations of $D_{\text{loc}}$, $D_{\text{nano}}$, $D_{\text{slow}}$ and $D_S$ for Aquivion, \textcolor{black} {as a function of the hydration-dependent structural variable $d_w$}. We find that, for both fast and slow species, diffusion coefficients steeply increase for $d_w \le 10$, followed by a milder dynamical enhancement at larger scales (water contents). \textcolor{black}{The local-scale diffusion coefficient of slow protons is an order of magnitude lower than than of fast protons at low hydrations, and reduced by a factor 5 at high hydrations. The diffusion of fast protons is significantly slowed-down at the nanoscale with respect to local scale, $D_{\text{nano}}$ being reduced by a factor 4 ($<$ 2) at low (high) hydrations with respect to $D_{\text{loc}}$. Moreover, nanoscale and mesoscale diffusion coefficients are comparable in the highly hydrated membranes, but $D_{\text{nano}}$ deviates from $D_S$ at $d_w \le 10$.} All-together these data indicate that transport in Aquivion is certainly scale-dependent, with a reduction of mobility occurring at all hydrations but with different strengths at molecular, nanoscopic and mesoscopic length scales. 
\subsection*{Anomalous water (sub-)diffusion}
\label{subsect:subdiffusion}
\begin{figure*}[t]
\centering
\includegraphics[width=1\linewidth]{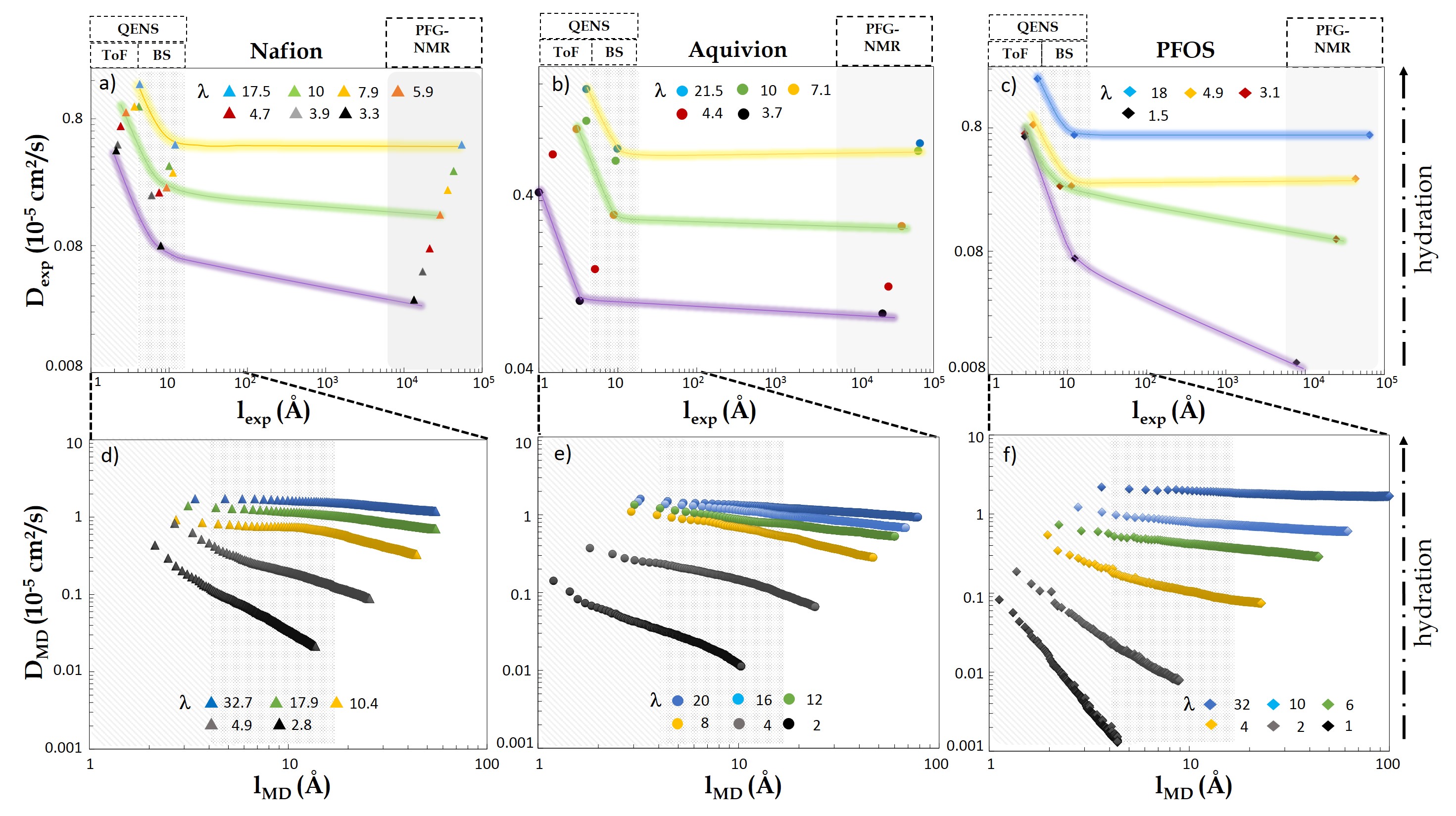}
\caption{Comparison of experimental (top) and numerical (bottom) effective diffusion coefficients of water, as a function of the characteristic length-scales denoted $l_{\text{exp}}$ and $l_{\text{MD}}$, respectively. The experimental values, $D_{\text{exp}}$, at the indicated values of $\lambda$ were obtained by QENS at local ($D_\text{loc}$) and nano- ($D_\text{nano}$) scales, and at microscopic scales by PFG-NMR ($D_S$) (see Figure~\ref{fig:3}). Data sets obtained for Nafion~\cite{Perrin2007} and Aquivion (this work) are shown, together with those pertaining to PFOS.~\cite{Berrod2015qens,Berrod2015} The determination of $l_{\text{exp}}$ is discussed in the main text. The numerical values $D_{\text{MD}}$ are the effective diffusion coefficients determined via Eqs.~(\ref{eq:msd}) and~(\ref{eq:Deff}), plotted parametrically against $l_{\text{MD}}(t)=\langle r^2(t)\rangle^{1/2}$.  Extended similar ranges of hydration $\lambda$ were probed in experiments and simulation.}
\label{fig:4}
\end{figure*}

The above picture, based on length-scales-dependent values of $D$, efficiently highlights the complex interplay between the structure of the confining environment and the transport of the adsorbed fluid. An important drawback, however, is that one is induced to think about the overall dynamics as the convolution of different mechanisms, active at different scales in a "piece-wise" Fickian-like fashion. This procedure somehow fails in appreciating the continuous character of the dynamics and, as a consequence, in fully characterizing the true {\em nature} of this latter. We now propose an alternative scenario which supplies unprecedented depth to the insight coming from spectroscopy. \textcolor{black}{In Figure~\ref{fig:4} we represent the experimental data discussed above in an uncommon fashion , where the fast protons diffusion coefficients are plotted as a function of the associated length-scales, $l_{\text{exp}}$}. This is, we believe, the most relevant result and take-home message of this work. Note that the representation of Figure~\ref{fig:4} is intrinsically different from that of \textcolor{black} {Figure~\ref{fig:3}}: $l_{\text{exp}}$ is now an attribute of the fluid motion itself, and ultimately depends on the dynamical features of the probe. $d_w$, in contrast, is a descriptor of the material degree of phase-separation. In the case of the QENS data, $l_{\text{exp}}^{\text{loc}}$ \textcolor{black} {can be associated to} the average size of the dynamical confinement ($2\sigma$, see above), while \textcolor{black}{we can define} $l_{\text{exp}}^{\text{nano}}$ as the inter-droplet distance . In the PFG-NMR measurements, $l_{\text{exp}}$ corresponds to the square-root of the mean-squared-displacement defined by the echo sequences ($l_{\text{exp}}^s=\langle r^2 \rangle^{1/2}=(6 D_S t)^{1/2}$). The values of $l_{\text{exp}}$ are reported in Table~S3$\dag$ for all cases. 

In the top panels of Figure~\ref{fig:4}, we plot on a log-log scale $D_{\text{loc}}$, $D_{\text{nano}}$, and $D_S$ measured in Nafion~\cite{Perrin2007}, Aquivion (same data of Figure~\ref{fig:3}), and PFOS~\cite{Berrod2015qens,Berrod2015} against the associated length scales, $l_{\text{exp}}$. The shaded lines interpolating the data at different $l_{\text{exp}}$ at a few values of $\lambda$ are convenient guides for the eye. At all values of $\lambda$, the dynamics monotonously slows-down by increasing $l_{\text{exp}}$, with a stronger variation at the smaller scales, and less pronounced at the larger. At the highest values of $\lambda$ the diffusion rate even saturate and, as a consequence, QENS and NMR measure diffusion coefficients with very similar values. In this view, the different experimental techniques hence seem to probe a {\em unique} scale-dependent dynamics, active with similar features in all the investigated PFSA materials. The dynamical ranges associated to the different experimental probes are, unfortunately, limited and exceedingly spread apart on the relevant time window, making questionable the continuous description suggested by the shaded lines. This issue should be lifted by considering extended sets of data, to fill the gap between the nm and $\mu$m scales. Alternative techniques like Neutron spin-echo or NMR relaxometry, however, also have limitations that hinder this possibility. (Neutron spin-echo techniques could be a viable option in this sense, as they allow to probe longer time-scales and possibly larger mass volumes. Unfortunately, it has been shown that this technique probes chain dynamics rather than fluid motions in PFSAs.~\cite{page2014nse} NMR relaxometry is the only technique providing information at the meso-scale and is very sensitive to polymer-water interactions. This technique indeed allows to characterize the dimensionality/topology of confinement, and was shown to be well suited for PFSA membranes.\cite{perrin2006relaxo} It does not allow, however, for a definition of a $Q$-vector and, consequently, for spatial resolution. Models including structural sizes and shapes can be used to analyze the obtained relaxation profiles, but determining values for the diffusion coefficients seems hazardous.) In our investigation we clearly hit the limits of application of even the most effective modern spectroscopy tools.

MD simulation can help in overcoming this gap. Indeed, the numerical integration of the equations of motion of the single molecules allows to evaluate at the microscopic level the correlation functions associated to the transport coefficients by simple forms of the fluctuation-dissipation theorem. This is in contrast with the above experimental measurements, which involve intricated modeling of very complex objects like the $S(Q,\omega)$. We have therefore performed an analysis similar as the above on the synthetic structures of Nafion, Aquivion and PFOS, generated on very extended time scales. Now, however, the starting point of the analysis is the mean-squared displacement of water molecules that can be written at the atomic level in the very simple form,
\begin{equation}
\langle r^2(t) \rangle = \frac{1}{N_w}\sum_{i=1}^{N_w} \left | \mathbf{r}_i(t) - \mathbf{r}_i(0) \right |^2.
\label{eq:msd}
\end{equation}
Here $N_w$ is the number of water molecules, and $\mathbf{r}_i(t)$ is the position vector of the oxygen atom of molecule $i$ at time $t$. Our data are shown in Figure~S12$\dag$ for Aquivion and Nafion , and in Ref.~\cite{Hanot2016} for the PFOS.

Beyond the expected feature that by decreasing $\lambda$, the mean-squared displacement at a given time decreases, we observe that the data fall on straight lines of slopes which progressively decrease with the hydration. At the highest value of $\lambda$ this slope is one, as expected for the (Fickian) case of bulk water. By lowering $\lambda$, however, the slopes decrease, indicating a power-law behavior $\langle r^2(t) \rangle\propto t^{\alpha}$ with $0<\alpha=\alpha(\lambda)< 1$, which is a signature of {\em sub-diffusion}.~\cite{metzler2014anomalous} This is a feature which has been substantially overlooked in previous numerical studies, with the exception of the observations reported in Ref.~\cite{savage2014persistent} and the characterization included in our Ref.~\cite{Hanot2016}. We demonstrate here, for the first time to the best of our knowledge, that it must be included in a complete rationalization of the experimental data.

In the presence of anomalous dynamics, one must consider a generalized form of the Einstein relation which connects the mean-squared displacement to the diffusion coefficient. We therefore define a time and length-scales dependent effective diffusion coefficient, $D_{\text{MD}}$, from the local slope of the mean-squared displacement as~\cite{allen1989computer}
\begin{equation}
\partial_t \langle r^2(t)\rangle = \frac{\alpha}{t}\langle r^2(t)\rangle = 6 D_{\text{MD}}.
\label{eq:Deff}
\end{equation}
Similarly to the representation of experimental data, in the bottom panels of Figure~\ref{fig:4} we plot $D_{\text{MD}}$ at different times $t$ as a function of the corresponding values of the observation length scale $l_{\text{MD}}(t)=\langle r^2(t)\rangle^{1/2}$, for all considered systems. (This representation is a slightly modified parametric representation of Eq.~(\ref{eq:Deff}).) The similarity of these data with the experimental counterpart of the top panels is \textcolor{black} {striking}, with the important difference that now the numerical analogous of $l_{\text{exp}}$ is a {\em continuous} function of time, encompassing the entire computationally explored time window and extending for a decade in the "no man's land" experimental region. At high $\lambda$, all PFSA materials are characterized by an almost constant $D_{\text{MD}}$, due to a water mobility dominated by the bulk-like diffusion in large pools of water. Decreasing water content is associated to important topological modifications that induce a drop of the diffusion coefficient at large $l_{\text{MD}}$ values. At given $\lambda$, the degree of diffusivity within the three systems is similar, as also seen from the values of the power law exponents and comparable values of the $D_{\text{MD}}$.

Experimental and numerical data therefore all together provide a convincing justification for the different values measured experimentally for $D_{\text{loc}}$, $D_{\text{nano}}$, and $D_S$, at any hydration level. These must {\em not} be ascribed to experimental resolution limitations, nor to selective sensitivity of the various techniques to different mechanisms, but are the manifestation of the true nature of the molecular dynamics in the materials, which is anomalous (non-Fickian) sub-diffusion. Neutron scattering and NMR techniques hence simply allow to sample at atomic, nanoscopic and macroscopic length scales the values of an effective diffusion coefficient, defined at the atomic level by Eq.~(\ref{eq:Deff}), without any need to invoke different types of motions and/or environment features.

In the case of ionic surfactants, it is also possible to clarify numerically the {\em origin} of the observed anomalous behavior.~\cite{Hanot2016} This is the result of a strongly heterogeneous and space-dependent dynamics characterizing water molecules lying at the interfaces compared to those further from the confining matrix, of bulk-like character. The observed {\em average} anomalous dynamics turns out to be ultimately related to an exchange mechanism between these dynamical groups, whose details depend on the water content and control the observed variable degree of sub-diffusivity. This analysis relies on the possibility of identifying the single water molecule position relative to the ionic channel boundaries, an option viable in simulation for the well-ordered phases formed by the PFOS and, more involved, for the ionomers. Additional implications of a space-dependent transport are highlighted below.
\begin{figure*}[t]
\centering
\includegraphics[width=1\linewidth]{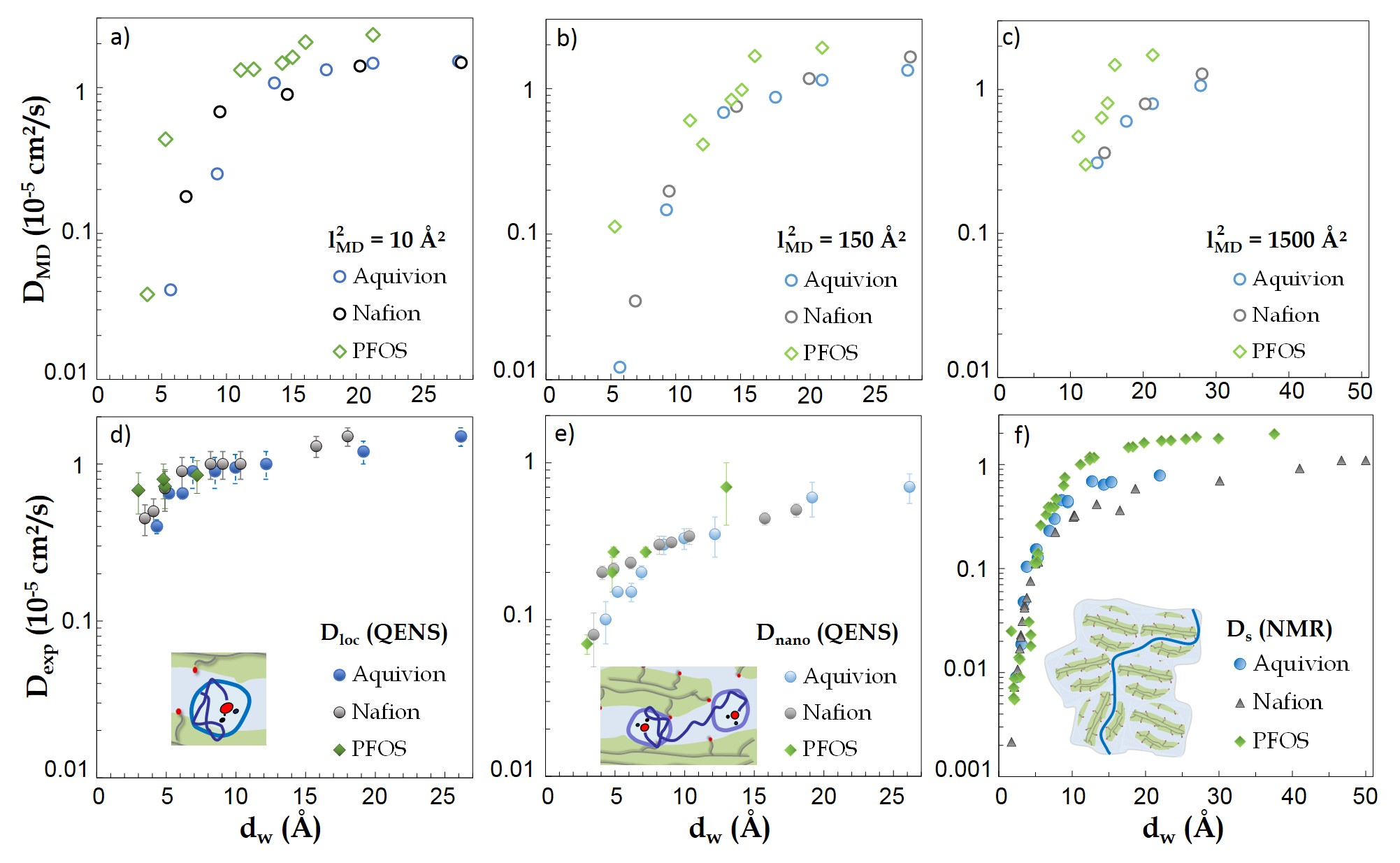}
\caption{
Diffusion coefficients of water molecules obtained by numerical simulation for Aquivion, Nafion, and PFOS \textcolor{black}{versus $d_w$ of Figure~\ref{fig:2}}. Values were determined by selecting the values of the numerical $D_{\text{MD}}$ of Figure~\ref{fig:4} at the length scales $(l_{\text{MD}}^*)^2\simeq$~10, 100 and 1500 \AA$^2$ (panels a), b), and c), respectively), for all materials in the investigated hydration ranges. These data can be directly compared to those shown in Figure~\ref{fig:3}, as discussed in the main text. {\em d)} Local $D_{\text{loc}}$ and {\em e)} nano-scale $D_{\text{nano}}$ diffusion coefficients determined by QENS using the generalized Gaussian model for localized translational motions.~\cite{Volino2006} We show the data for Aquivion, Nafion,~\cite{Perrin2007} and surfactants.~\cite{Berrod2015qens,Lyonnard2010} {\em f)} Water self-diffusion coefficient, $D_S$, measured by PFG-NMR of Aquivion, Nafion \cite{Perrin2007} and PFOS surfactant.\cite{Berrod2015} The diffusion coefficients are plotted as a function of $d_w$ (see Figure~\ref{fig:2}), to highlight the complex interplay between the structure of the confining environment and the dynamical behavior of the adsorbed fluid. All details of the data analysis can be found in the ESI$\dag$.}
\label{fig:5}
\end{figure*}
\subsection{Generic Dynamical Behavior in PFSAs}
\label{subsect:generic behavior}
The above conclusions can be further strengthened by selecting the values of the numerical $D_{\text{MD}}$ corresponding to $l_{\text{MD}}^*\simeq l_{\text{exp}}^{\text{loc}}$, $l_{\text{exp}}^{\text{nano}}$, and $l_{\text{exp}}^{\text{s}}$ at each $\lambda$, for all materials. (We note that our simulation box size is typically of the order of a few nanometers only, which is far from the $\mu$m range probed by NMR. We have chosen, however, to consider a value $l_{\text{MD}}^*\simeq 100$~\AA$\,$ as the one appropriate to reproduce the corresponding experimental values, at least at the higher values of $\lambda$. Indeed, as previously shown for the case of surfactants,~\cite{Hanot2016} we have found that in high hydration conditions the values of $D_{\text{MD}}$ are length-scale independent for $l>l_{\text{MD}}^*$.) \textcolor{black}{The MD data are shown in Figures~\ref{fig:5}~\textcolor{black}{a), b), c)} versus $d_w$, and can be directly compared to the experimental results $D_{\text{loc}}$, $D_{\text{nano}}$ and $D_S$ obtained on Aquivion, Nafion and PFOS (Figures~\ref{fig:5}~ d), e), f). Interesting generic features of the dynamics in PFSAs can be emphasized by such collective comparison.} In Figure~\ref{fig:5}~d) and e) we observe values of experimental $D_{\text{loc}}$ and $D_{\text{nano}}$ of Nafion and Aquivion which superimpose in the limit of the error bars, with a diffusion which steeply increases for $d_w\le 10$ \AA, followed by a more moderate enhancement at larger water contents. No significant difference is therefore found in the two cases of long and short side-chain materials. Surprisingly, however, we find a very similar behavior also in surfactants (considering a multi-Lorentzian approach,~\cite{Berrod2015qens} or the Gaussian model treatment~\cite{Lyonnard2010}). We are therefore in the presence of a robust generic dynamical behavior, common to all investigated PFSA systems, independently on the details of the macro-molecules structures or even the presence of an hydrophobic backbone. We have verified that while the precise values of the dynamical parameters indeed depend on the particular model used to represent the data, the generic features of the dynamics do not. We are therefore convinced that the above picture is general and qualitatively model-independent.

The experimental data for $D_S$ for all materials shown in Figure~\ref{fig:5}~f) deserve some additional comment. As already noted in Ref.~\cite{Berrod2015}, at small $d_w$ all materials exhibit, again, very similar behavior upon increasing water content. At high $d_w$, in contrast, a distinctive behavior is observed, due to significant differences in the total volume fraction of water at constant $d_w$ in the different materials. We can rationalize this discrepancy by noting that when hydrophobic aggregates are diluted in water-filled domains (see snapshots in Figure~\ref{fig:1}), the structural feature that ultimately controls the dynamics of the adsorbed fluid is simply the presence of obstacles, whatever their size or shape.~\cite{Berrod2015} Indeed, if we represent $D_S$ as a function of $\Phi$ (Figure~S11$\dag$), data of all PFSA materials fall on the same curve at large water uptake, typically for $\Phi \geq 0.2$. We therefore conclude that a substantial invariance of the dynamical behavior of the absorbed water under changes of the details of the macro-molecular structure survives at all length-scales, in every case where confinement at the nano-scale plays a non-negligible role.

\textcolor{black} {Despite the relative crudeness of the molecular models used in our simulations, we surprisingly find a remarkable agreement between the experimental and numerical data sets, both showing a continuous reduction of mobility while increasing the length scale at a given hydration, this effect being much more pronounced at low hydration.} The MD data independently confirm that Nafion and Aquivion membranes are characterized by very similar modifications of the transport properties upon hydration. These modifications are general for the entire class of sulfonated materials considered here, irrespective to the detailed molecular structures. This is clearly demonstrated by the data pertaining to PFOS, which superimpose to those referring to the much more structured ionomers, in complete agreement with the experiments.
\begin{figure*}[t]
\centering
\includegraphics[width=0.8\textwidth]{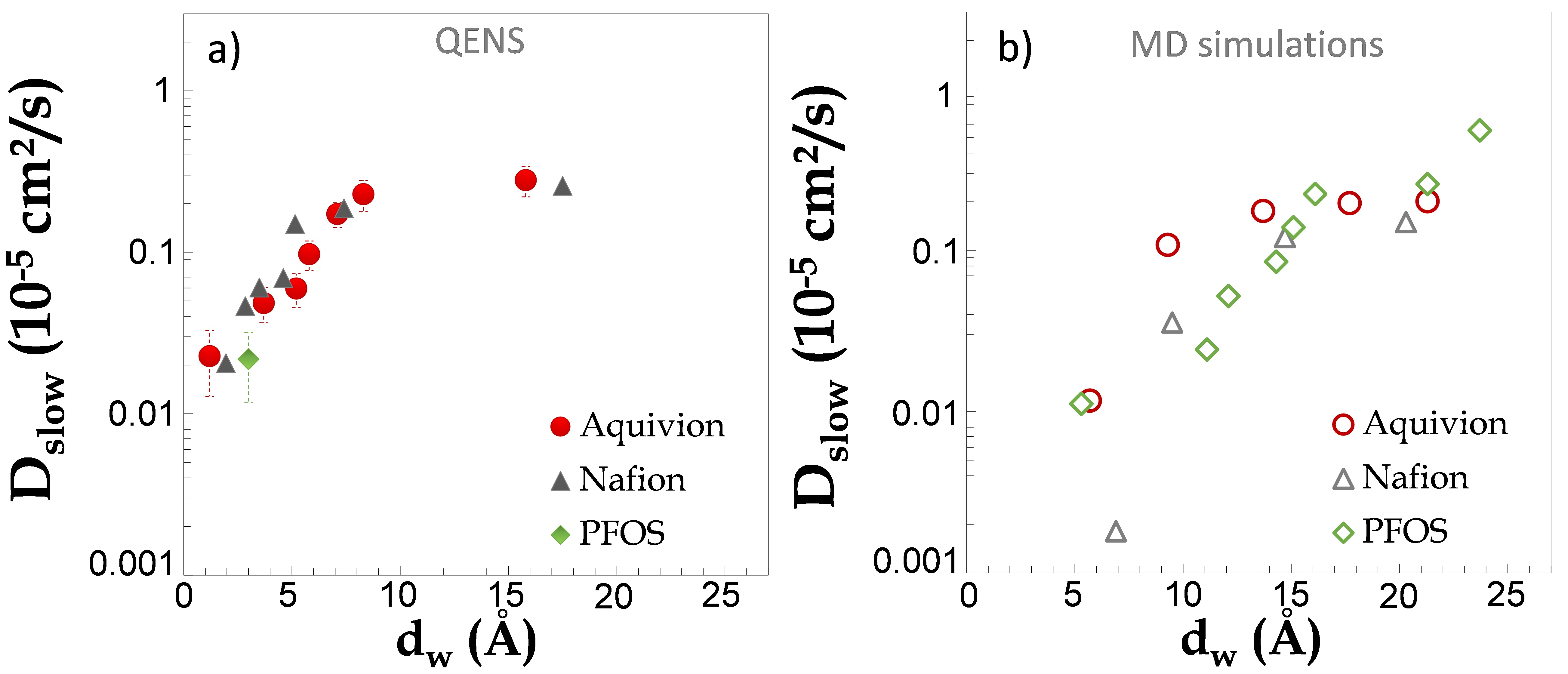}
\caption{ {\em a)} Experimental slow diffusion coefficients for Nafion~\cite{Perrin2007} PFOS and Aquivion. These data are determined from the analysis of the slow component of the spectra, $D_{\text{slow}}=\sigma_s^2/\tau_s$, as discussed in the main text and detailed in the ESI$\dag$. $d_w$ is the typical size of the ionic domains (see Figure~\ref{fig:2}). {\em b)} Diffusion coefficients of the hydronium ions, obtained by numerical simulations for all materials. Data have been determined by using an analysis analogous to that for water (see Figure~\ref{fig:5}), with $(l_{\text{MD}}^*)^2\simeq$ 10 \AA$^2$.
}
\label{fig:6}
\end{figure*}
\subsection{Transport of the hydronium ion}
\label{subsect:hydronium diffusion}
The above data analysis, \textcolor{black} {integrating all numerical and experimental information obtained for the three investigated PFSA compounds,} also delivers sensible information about proton conduction. Full ionization is observed in PFSA materials already at very low water content, due to the super-acidic character of sulfonic acid. In particular, it has been shown that hydronium ions are formed in Nafion as soon as just one water molecule per ionic group is loaded in the system.~\cite{dalla2014IR} Proton conduction is therefore certainly strongly related to the mobility of the \ce{H3O+} ions, which in turn, depends on the dynamics of the surrounding water molecules discussed above. The organization of a well-connected hydrogen-bonds network in the ionic phase, however, is also critical, via cooperative formation and annihilation of the \ce{H}-bonds in the hydration shell of the ion. Hydrated species as Zundel and Eigen-ions are probably formed,~\cite{kreuer2000complexity,kreuer2004transp-rev} and contribute to promote charge transfer across the material. 

The relevance of these two mechanisms, makes the experimental quantification of proton transfer difficult. On one side, the proton diffusion coefficient is often derived from conductivity measurements via the Nernst-Einstein relation, although this is strictly valid in the case of diluted solutions only, where diffusion of charge carriers and proton diffusion are identical processes.~\cite{schuster2008proton} On the other, QENS and NMR fail to discriminate the behavior of charge carriers from that of hydration protons. Classical (non-reactive) MD simulation as that we employ here, obviously distinguishes between the \ce{H3O+} and \ce{H2O} species but, unfortunately, cannot account for the structural diffusion controlled by the \ce{H}-bond dynamics. Advanced effective techniques like the Empirical Valence Bond approaches~\cite{voth2012evb,savage2014persistent,arntsen2016simulation} have been developed to lift this limitation, but a convincing complete modeling of proton transport comprising the plain complexity of the PFSA environment is still not available. 

Interestingly, it turns out that even if suppressing structural proton conductivity, our MD simulation can be quite specific on the vehicular aspect, encoded in the \ce{H3O+} molecular dynamics. In particular, we can now discuss the nature of the slow protons population characterized by $D_{\text{slow}}$ (see Figure~\ref{fig:3} for the Aquivion), a point we left open in the discussion above. Based on our analysis of QENS data (see ESI$\dag$), the dynamics of the slow proton can be described as hopping in a very restricted environment ($\sigma_s$) of a few \AA, on typical timescales ($\tau_s$) of hundreds of ps. On increasing hydration, hopping is faster on larger distances but the resulting motion remains highly localized on the time-scale of the experiments, below the ns scale. This is in contrast with the diffusive motion of fast protons, which are most probably located well within the fluid channels, far from the phase boundaries, and are capable to move over nano-metric distances, on ns time scales.

This picture is agnostic about the {\em identity} of the slow protons, an issue we can attack by simulation. In Figure~S13$\dag$, we plot $\langle r^2(t)\rangle$ calculated via the analogous of Eq.~(\ref{eq:msd}) for the hydronium ions, in Aquivion and Nafion at the indicated values of $\lambda$. Similar results are obtained for the PFOS. A variable degree of sub-diffusion is clear in these data, with power-law exponents at low $\lambda$ even lower ($\alpha\simeq 0.3$) than those pertaining to water molecules in the same hydration conditions. The anomalous character of the dynamics therefore seems to be stronger for the charged species, a feature one can correlate to the enhanced ionic interactions with the acid functions decorating the ionic domains boundaries. (We note, for completeness, that this is at odds with Refs.~\cite{song2015nanometer,mcdaniel2016importance}) From these data we can extract time and length scales-dependent effective diffusion coefficients based on Eq.~(\ref{eq:Deff}), which we now sample at values of $l^*_{\text{MD}}\simeq l_{\text{exp}}^{\text{slow}}=2\sigma_s$, in complete analogy with the case of water molecules. We plot these data for all materials in Figure~\ref{fig:6}~b), as a function of $d_w$, \textcolor{black}{and compare them to the experimental} values of $D_{\text{slow}}=\sigma_s^2/\tau_s$ shown in panel a), at the corresponding $d_w$. The agreement between the QENS diffusion of slow protons and the MD diffusion of the hydronium ions is quite remarkable, in terms of both absolute values and modifications with the hydration state. 

\textcolor{black}{As observed above, the number of slow protons is found to be on average three per ionic groups, independently of the hydration level (Figure S8$\dag$). In particular, at $\lambda=1$, once one hydronium ion is formed, all three protons feature slow hopping motions, with no established capability of producing long-range diffusivity. These results can be rationalized by assuming that slow protons are likely to be in strong interaction with the sulfonic groups, and are therefore mostly condensed at the hydrophilic/hydrophobic interfaces. Note that the interfacial region is not dramatically modified even when significantly increasing the total amount of adsorbed water, as its spatial expansion is determined by local ionic interactions among the sulfonate groups and the first layer of the surrounding molecules. For instance, we have found that in ionic surfactants the fraction of hydronium ions localized in the first coordination shell of the \ce{SO3-} groups varies from  $95\%$ to $50\%$ at $\lambda=2$ and 32, respectively (see Figure~4 of Ref.~\cite{Hanot2015} for PFOS). Similar results are also reported for the fully atomistic simulation of the membrane of Ref.~\cite{liu2010ew}, and the presence of long-lived hydronium ions at the Nafion interface was also demonstrated in Ref.~\cite{devanathan2007}, even at very high $\lambda$.}

\textcolor{black}{On this basis, we can therefore reasonably assume that the slow protons can pertain (at least in part) to hydronium ions, in particular at low $\lambda$. As a consequence, we can associate the slow relaxation mechanisms revealed by the QENS analysis mainly to the dynamics of \ce{H3O+} moieties, which are significantly adsorbed at the interfacial layers in all hydration conditions. We complement this conclusion with two caveats. First, we note that at high water content, the interfacial layer most probably contains additional water molecules slowed down by the interaction with the charged interfaces, and intercalated between the sulfonated groups and the protonic species. Second, hydronium ions immersed in the water domains (see systems snapshots in high hydration conditions in Figure~\ref{fig:1}) are likely to behave differently compared to the "interfacial" hydroniums. They also contribute, however, to the average MD hydronium ion diffusion coefficients plotted in Figure~\ref{fig:6} b), with a strength that we cannot quantify at present. A space-dependent analysis of the mean square displacements across the ionic channels, similar to that we performed in Ref.~\cite{Hanot2016} for water absorbed in PFOS, could help in further describing the details of ion motions. These could include, for instance, the occurrence of constrained rotations due to the presence of a first water coordination sphere with reduced mobility, a feature which cannot be excluded on the basis of the present data analysis.}
\section{Conclusion}
\label{sect:conclusions}
Confinement of fluids and charges at sub-nanometric scales is at the heart of many energy applications. It can induce spectacular enhancement of performances in some cases, like in storage devices,~\cite{chmiola2006anomalous} or in blue energy applications based on osmotic power harnessing.~\cite{siria2013giant} It can be detrimental in others, as it is the case in low hydration conditions of the perfluoro-sulfonic acids compounds we have investigated here. It is obviously crucial to grasp the general physical mechanisms laying at the bottom of this behavior, which cannot depend on the details of the macro-molecular structure of the materials, but only on a few most fundamental attributes. As a consequence, we have chosen to test the statistical significance of our conclusions by considering and analyzing simultaneously very complex ionomers, like Aquivion and Nafion, and simpler "model" materials as perfluoro-sulfonated ionic surfactants. We have therefore combined in a comprehensive study an extremely large corpus of data sets related to these materials and produced by means of advanced spectroscopy techniques, together with the results of extensive MD simulations of synthetic versions of the same systems. 

By treating at the same level all the information, we have contributed solid insight on the true nature of the dynamics of the chemical moieties adsorbed in the ionomer charged hydrophobic matrix. We have demonstrated, in particular, that all materials share an anomalous, sub-diffusive, dynamical character, never discussed before into such details. This finding can be rationalized in terms of highly heterogeneous transport properties, due to the complicated interplay between spatial constraints (confinement at the nanoscale) and direct interactions with the interfaces. The picture we propose here must be taken into account for a complete rationalization of the experimental data and, we believe, will be beneficial for any future advance in the rational design of new proton-conducting ionomers or composites for fuel cells applications. The approach we have followed establishes correspondences with advanced topics in statistical mechanics, resulting in unprecedented possibilities for even more detailed analysis of the physics encoded in Neutron scattering data. This work also indicates interesting connections and possibilities of cross-fertilization with other research fields, including biophysics.

\section{MATERIALS AND METHODS}
\label{sect:methods}
Our analysis is based on numerous data sets, stemming from experiments or generated numerically by MD simulation. A few details on the experimental and numerical techniques, and on protocols followed for the measurements performed are given below. More extended information about experimental raw data, modeling and fitting procedures, and values of the obtained parameters are included in the ESI$\dag$. New unpublished data and analysis are reported in this work, and combined to previous results in an original integrated general picture of the behavior of water and charged adsorbed in PFSA compounds. More in details, our discussion will be based on the following data-sets:

\noindent
{\bf DS1. Aquivion--} {\em This work:} Small-Angle X-Rays Scattering (SAXS), Pulsed Field Gradient Nuclear Magnetic Resonance (PFG-NMR), multi-resolution Quasi Elastic Neutron Scattering (QENS) (data analyzed with a multi-population/multi-scale model,~\cite{Volino2006} details are reported in the ESI$\dag$), MD. {\em Previous data:} QENS.~\cite{Berrod2015qens}

\noindent
{\bf DS2. Nafion--} {\em This work:} MD. {\em Previous data:} SANS,~\cite{Perrin2007} NMR,~\cite{Perrin2007,Berrod2015} multi-resolution QENS.~\cite{Perrin2007}

\noindent
{\bf DS3. PFOS--} {\em Previous data:} SAXS,~\cite{Lyonnard2010,Berrod2015} NMR,~\cite{Berrod2015} QENS~\cite{Berrod2015qens,Lyonnard2010} and MD.~\cite{Hanot2015,Hanot2016}
\subsection{Materials}
\label{subsect:materials}
50~$\mu$m-thick Aquivion membranes of equivalent weight $EW=$ 790 and 850 $g/eq$ were purchased from Solvay. For full acidification, the as-received membranes were first acidified in $1M$ nitric acid (60$^o$~C for 2 hours) and subsequently thoroughly rinsed in pure water (60$^o$~C, 2 hours, 2 repetitions). We measured the sorption isotherms (shown in Figure~S1$\dag$) to convert the relative humidity (RH) into the water content. This is expressed in terms of the macroscopic water volume fraction, $\Phi$, or the local variable $\lambda$, defined as the number of water molecules per sulfonic acid group. Samples were prepared at relative humidities ranging from 11\% to 100\%~RH, providing values of $\lambda$ in the range 2 to 20 (Figure~S1 and Tab.~S1-S2$\dag$). Nafion and Aquivion samples generated by simulation were initialized in similar hydration ranges.

\subsection{Methods}
\label{subsect:methods}

{\bf 1. SAXS}. Measurements were performed at the European Synchrotron Radiation Facility (Grenoble, France) on the BM02-D2AM beam-line, with an incoming X-rays energy of 14~keV. Aquivion membranes (850~g/eq) were first equilibrated for several hours under various relative humidity conditions and then enclosed in hermetic cells (mica windows). The water vapor pressures were imposed by using saturated salt solutions. Standard procedures for background subtraction, data corrections, and radial averaging were applied to obtain the final 1-dim $I(Q)$ scattering spectra (Figure~S2$\dag$), with $Q$ the exchanged wave-vector.

{\bf 2. PFG-NMR}.
The water self-diffusion coefficient, $D_S$, was determined as a function of the hydration level using the PFG-NMR technique. The measurements were performed at room temperature with a low-field analyzer Bruker Minispec, operating at 20~MHz for the $^1$H nucleus, allowing a magnetic field gradient up to 4~$T m^{-1}$. A spin-echo sequence was used and $D_S$ obtained from $E(G)/E(G=0)=\exp\left(-(\gamma G\delta)^2 D_S(\Delta-\delta/3)\right)$. Here, $\gamma$ is the gyromagnetic ratio of the investigated nucleus ($\gamma=2.6752\times 10^4$~rad s$^{-1}$~$T^{-1}$ for $^1$H), $G$ the intensity of the magnetic field gradient, and $\delta$ the duration of the gradient pulse. The mean-squared displacements of water molecules in the Aquivion membranes (790~g/eq) were varied between 1 and 50~$\mu m^2$, depending on both $D_S$ and the diffusion time $\Delta$ (ranging from 10 to 20~ms).

{\bf 3. QENS}. The collection of QENS spectra on extended timescales was realized by combining multi-resolution time-of-flight (ToF) and high resolution backscattering (BS) experiments. The ToF experiments were performed on the spectrometers Mib\'emol at the Laboratoire L\'eon Brillouin (LLB), Saclay, France, and IN5 at the Institut Laue Langevin (ILL), Grenoble, France. Incoming wavelengths $\lambda_{\text{inc}}=$ 8 and 5~\AA ~were used on IN5,  $\lambda_{\text{inc}}=$ 9 and 6~\AA ~on Mib\'emol, providing elastic resolutions of 20, 90, 30 and 100~$\mu eV$, respectively. Backscattering experiments were carried out at the ILL (IN16), with $\lambda_{\text{inc}}$ = 6.27~\AA ~and a resolution of 1~$\mu$eV. Strips of Aquivion membranes (790~g/eq, 3$\times$11 cm$^2$) were enclosed in a cylindrical cell containing saturated salt solutions, setting the RH in the 11 to 100~\% range. To minimize multiple-scattering effects, the sample transmissions were verified to be higher than 95~\%. The quasi-elastic dynamic structure factor spectra, $S(Q,\omega)$, at energy transfer $\omega$ were obtained using the standard ILL and LLB routines, and corrected to account for the presence of the sample container and the efficiency of the detector. A flat vanadium sample was used to measure the experimental resolution function. 

Data have been treated with the QENSH software (available at: www-llb.cea.fr/en/Phocea/Page/index.php?id=21) and analyzed following the method we used in Ref.~\cite{Perrin2007} to clarify the dynamics of protons in Nafion membranes. A large number of data sets recorded on Nafion, in hydration conditions ranging from the dried to the fully swollen states, were best accounted for by employing several quasi-elastic components, due to the existence of protons with different unusual dynamical features when confined in soft charged media. Localized motions were analyzed using a standard single-Lorentzian approach, diffusive motions in terms of the generalized Gaussian model for localized translational motion (GMLTM).~\cite{Volino2006} The generalized GMLTM was developed to describe the diffusion of species confined in soft ill-defined environments (e.g. droplets) with no impermeable boundaries, and includes the possibility of both intra- and inter-droplet diffusion. Here we have used a similar procedure to treat the $S(Q,\omega)$ of Aquivion. Details on the generalized GMLTM, the procedure to fit consistently and simultaneously the multi-resolution data and the obtained dynamical parameters are included in the ESI$\dag$.

{\bf 4. MD simulation}. Simulations have been conducted with LAMMPS~\cite{plimpton1995fast} by using the macro-molecular representation and force-field introduced in Ref.~\cite{Hanot2015} for hydrated ionic surfactant systems. In particular, we demonstrated that our re-optimized force-field (inspired to that of Ref.~\cite{allahyarov2008simulation}) correctly encodes the phase behavior and the main self-organized structural features of real PFOS at different hydration,~\cite{Hanot2015} and characterized water diffusion within the hydrophilic domains~\cite{Hanot2016}. Due to the already mentioned similarity of PFOS with the side-chain of Nafion, we employ that model as the primary building-block to upscale the molecular description to the entire ionomer, by grafting the amphiphilic units to a strongly hydrophobic polymer backbone. By tuning the size of the side-chains and the grafting repetition pattern we can now straightforwardly reproduce the structure of Nafion and Aquivion at the desired value of equivalent weight, respectively (see Figure~\ref{fig:1}). Similarly, a system formed by side-chains of tuned length only is our model for PFOS. We can thus count on consistent synthetic templates for the systems studied in the experiments, with a physically sound behavior on length scales ranging from local lamellar-like structures to mesoscopic ionic and hydrophobic domains.

More specifically, simulations were designed to allow for a quantitative analysis of the adsorbed species dynamics. We therefore employ well-established non-reactive all-atoms models for water (\ce{H2O}) molecules~\cite{berendsen1987missing} and hydronium (\ce{H3O+}) complexes~\cite{kusaka1998binary} (see Figure~\ref{fig:1}), with the screening of long-range interactions of Ref.~\cite{fennell2006ewald} We adopt, in contrast, a coarse-grained representation for the polymer backbone and charged amphiphiles, where one and two interaction centers (represented as beads in Figure~\ref{fig:1}) mimic entire \ce{CF2} and \ce{SO3-} groups, respectively.~\cite{Hanot2015} The number of \ce{H3O+} complexes in the simulation box is dictated by the number of \ce{SO3-} groups (for charge neutrality), the number of water molecules per \ce{SO3-} fixes the hydration $\lambda$. Details of the interaction potential and simulation methods can be found in Ref.~\cite{Hanot2015} Snapshots of the obtained morphologies for all materials are shown in Figure~\ref{fig:1} at low ($\Phi\simeq$ 0.1) and high ($\Phi\simeq$ 0.4) water volume fractions. The high resolution features of the snapshots, of course, depend on the details of the model employed. These structures, however, still give a quite precise idea of the typical morphologies one can expect in real systems, an information much harder to obtain experimentally.
\section{Acknowledgements}
\noindent
The author thank the ILL and LLB for beam-time allocation; B. Frick, J. Ollivier and J.-M. Zanotti for help with the QENS measurements and discussions. Q. Berrod thanks the Grenoble-Alpes University (UGA), and the ILL for funding. S. Hanot thanks the ILL for funding.
\bibliography{references}
\end{document}